# Prototyping and Evaluating a Real-time Neuro-Adaptive Virtual Reality Flight Training System

Evy van Weelden, Jos M. Prinsen, Caterina Ceccato, Ethel Pruss, Anita Vrins, Maryam Alimardani, Travis J. Wiltshire, and Max M. Louwerse

***Abstract*— Real-time adjustments to task difficulty during flight training are crucial for optimizing performance and managing pilot workload. This study evaluated the functionality of a pre-trained brain-computer interface (BCI) that adapts training difficulty based on real-time estimations of workload from brain signals. Specifically, an EEG-based neuro-adaptive training system was developed and tested in Virtual Reality (VR) flight simulations with military student pilots. The neuro-adaptive system was compared to a fixed sequence that progressively increased in difficulty, in terms of self-reported user engagement, workload, and simulator sickness (subjective measures), as well as flight performance (objective metric). Additionally, we explored the relationships between subjective workload and flight performance in the VR simulator for each condition. The experiments concluded with semi-structured interviews to elicit the pilots' experience with the neuro-adaptive prototype. Results revealed no significant differences between the adaptive and fixed sequence conditions in subjective measures or flight performance. In both conditions, flight performance decreased as subjective workload increased. The semi-structured interviews indicated that, upon briefing, the pilots preferred the neuro-adaptive VR training system over the system with a fixed sequence, although individual differences were observed in the perception of difficulty and the order of changes in difficulty. Even though this study shows performance does not change, BCI-based flight training systems hold the potential to provide a more personalized and varied training experience.**

## I. INTRODUCTION

PASSIVE brain-computer interfaces (pBCIs) can predict mental state(s) based on neural activity [1], [2], enabling personalized feedback accordingly to enhance user experience and learning. Wearable sensors such as electroencephalography (EEG) electrodes that measure brain activity can be used for the purpose of mental state monitoring [3], e.g., continuously inferring user's level of cognitive workload or task engagement.

The research reported in this study is funded by the MasterMinds project, part of the RegionDeal Mid- and West-Brabant, and is co-funded by the Ministry of Economic Affairs and Municipality of Tilburg awarded to MML. *(Corresponding author: E. van Weelden).*

Evy van Weelden is with the Royal Netherlands Aerospace Centre, Amsterdam, NL (e-mail: evy.van.weelden@nlr.nl). Jos M. Prinsen, Travis J. Wiltshire and Max M. Louwerse are with Tilburg University, Tilburg, NL (e-mail: j.m.prinsen@tilburguniversity.edu; t.j.wiltshire@tilburguniversity; m.m.louwerse@tilburguniversity.edu). Caterina Ceccato, Ethel Pruss, Anita Vrins and Maryam Alimardani are with Vrije Universiteit Amsterdam, Amsterdam, NL (e-mail: c.ceccato@vu.nl; e.pruss@vu.nl; a.m.vrins@vu.nl; m.alimardani@vu.nl).

This work involved human subjects in its research. Approval of all ethical and experimental procedures and protocols was granted by the Research Ethics Committee of Tilburg School of Humanities and Digital Sciences (Application No. REDC2021.36). This manuscript has Supplementary Materials.

Integrating these sensing technologies with training devices supports the development of neuro-adaptive training simulations, i.e., training simulations that dynamically adapt to a user's experience through the use of pBCIs. Virtual Reality (VR) technology further adds the advantages of being portable, immersive, and easily combinable with wearable sensors. In aviation training, the use of VR combined with EEG technology is expected to significantly improve the efficacy of simulation-based training, offering real-time feedback in an immersive and customizable training environment [3].

VR has already been widely used for training across different operational contexts, such as surgical training [4], maritime training [5], and aviation training [6], [7]. These fields, although merely examples, require operators to be able to handle a great task load. Recent advances in pBCIs allow for monitoring workload during (simulated) training, for the purpose of enhancing the individual's learning trajectory or keeping track of diminished states for learning, safety and performance, as demonstrated in [8]-[14].

In the aviation domain, it is relevant to monitor pilot workload for several reasons. First, monitoring workload is important because of its relation to pilot performance [15]-[18], and consequently flight safety. Secondly, low and high workload can also negatively impact operator training [19]. For optimal training, one needs to operate in a mental 'sweet spot' in-between an overload and underload [19]. In fact, Hart [19] recommended that critical situations should be practiced in this optimal state as well, at first – for learners to acquire the necessary skills and knowledge before stress-inducing situations are actually trained or encountered.

Real-time applications of pBCIs, that have the potential for detecting and maintaining an optimal workload state for learners, have only received limited attention, [12]-[14]. And, importantly, studies on such closed-loop pBCI systems – in which task or simulator adaptations are implemented without human interference – are even more limited, [20]. The underlying reason for this scarcity, is that there are numerous challenges involved in the development of such systems. For example, it is difficult to establish a ground truth of mental state(s) of interest [21], and even with (often lengthy) calibration sessions, it remains difficult for pBCI models to be applicable to all subjects [22].

Applying pBCI models to multiple subjects is a key challenge within the broader issue of transfer learning. In BCIs, *transfer learning* involves (successfully) applying models across different subjects, tasks, sessions and devices [23], [24], though this is often difficult to achieve or even infeasible. In our previous study [25], we addressed one aspect of the transfer learning problem by attempting to develop a subject-independent and calibration-free pBCI model for workload in VR flight training. In the current paper, we expand on our prior work by applying the same (pre-trained) subject-independent pBCI classifier to real-time flight training scenarios. We implemented a prototype of a real-time neuro-adaptive VR flight training system and explored pilots' subjective and behavioral responses when they were trained with such system. More specifically, we compared the pBCI-based neuro-adaptive training to a non-adaptive, fixed-order training in an experimental study with a group of novice military pilots. The purpose of this study is to evaluate whether pBCI paradigms can be useful in the training of novice pilots, and to give recommendations on how to improve and facilitate the application of pBCIs in (VR) simulated flight training.

## II. pBCI MODEL

### A. Model overview

The pBCI model implemented in the current study was previously reported in [25]. This machine learning (ML) model was trained to classify two levels of workload (Low vs. High), with the use of 32-channel EEG data from six novice military pilots. The pilots performed 12 trials of speed change task in VR simulated flights in a Pilatus PC-7 Turbo Trainer aircraft; in half of the trials, they only conducted the speed change task (Low workload) and in the other half, they had a secondary task (auditory *N*-back) in addition to speed change, which increased the overall task load (High workload).

Our approach to feature selection and classification was based on the architecture from [10]. Upon manual cleaning of the EEG data, we calculated spectral powers in alpha, beta, and theta frequency bands for all 32 channels in all trials. Additionally, for each channel, the EEG Engagement Index (beta/(alpha+theta)) [26] was computed as an additional feature. This produced a total of 128 features for each trial. EEG features were then selected using recursive feature elimination (RFE). The selected EEG features, ranked in order of importance, include "Beta P8", "Alpha P8", "Theta Oz", "EEG Engagement T8", "Beta F8", "Theta T7", "EEG Engagement P8", "Theta CP6", "Beta FP2", and "Theta FP1". These features are represented by their frequency bands (i.e., theta, alpha, beta, and the aforementioned EEG Engagement Index) and electrode locations (i.e., P = Parietal, O = Occipital, T = Temporal, F = Frontal, CP = Centroparietal, FP = Pre-frontal). For a map of EEG electrode locations, please refer to [25].

A stacking classifier was developed to predict the two levels of workload and consisted of an ensemble model of Support Vector Machine (SVM), Random Forest (RF) and Logistic Regression (LR) in the first layer, and SVM as second layer. Hyperparameters were optimized using grid search. The classifier was trained and tested over ten iterations with data split into 70% for training and 30% for testing in each iteration.

The ML model demonstrated robust performance, achieving a mean accuracy of .78 (*SD* = .07), a mean F1 score of .76 (*SD* = .09), a mean precision of .78 (*SD* = .11), and a mean recall of .76 (*SD* = .14).

### B. Real-time implementation

The real-time implementation of the neuro-adaptive VR flight training system was achieved using two main components: the VR simulator and an external computer. The external computer was connected to the EEG system and ran local Python code to classify workload based on input signals. A schematic overview of the system is provided in Fig. 1. A connection was made between the external computer and the simulator via an ethernet cable and modules were scripted to receive and send information. The simulator featured a graphical user interface (GUI) displayed on an external touchscreen monitor, allowing experimenters to set conditions and start or stop trials with button presses. Once the simulator started running a trial, the recording of the EEG data was initiated and continued for the duration of the task (120 seconds). After each trial, the simulator would automatically stop and reset the VR environment and flight coordinates to the start position, preparing for the next trial.

The MNE package for Python [27] was used to clean the signals in real-time using bandpass (3-35 Hz) and notch (50 Hz) filters, along with automatic independent component analysis (ICA). It was also utilized to compute spectral powers in the theta, alpha, and beta frequency bands across all 32 channels for the whole duration of recording (120 seconds). Next, the 10 EEG features necessary for classification of workload by the pre-trained model were selected and structured as input for the pBCI classifier. The classifier would then label the trial as 'Low' or 'High' workload.

Based on the classification output, the level of difficulty in the next trial was changed: when workload in one trial was classified to be 'High', the difficulty in the next trial decreased one level; when workload was classified to be 'Low', the difficulty of the next trial increases with one level. We argued that this would keep trainees in an optimal state of workload for training.

This closed-loop system was repeated for five iterations for the purpose of our experimental study. In cases that the classifier indicated a decrease in difficulty whilst the system was on the easiest level, or vice versa, the same level of difficulty was repeated in the following trial. Each new trial needed to be started manually in the GUI, which allowed some time for a break and questionnaires.

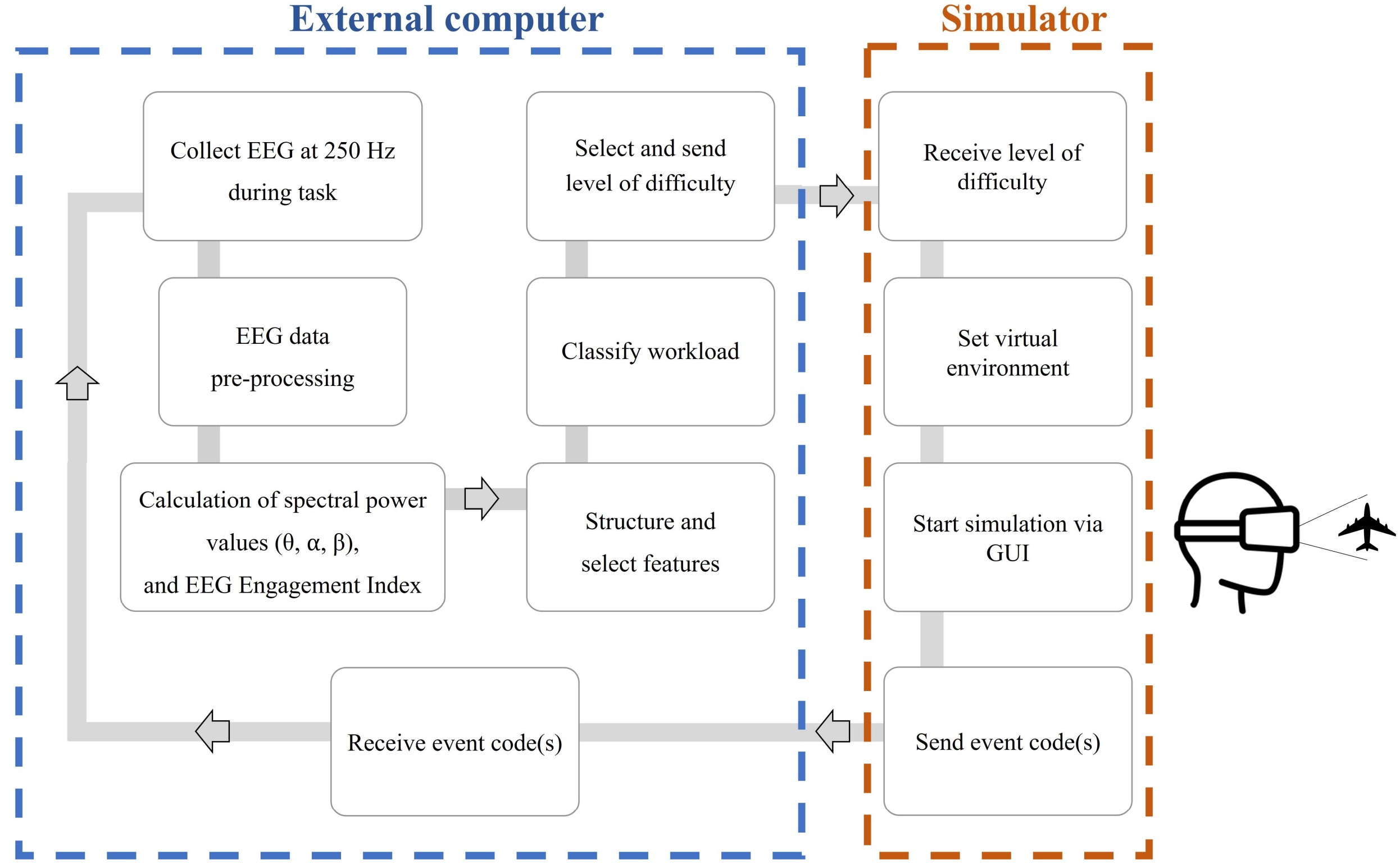

**Fig. 1.** Schematic pipeline of the neuro-adaptive flight training system. This pipeline was run in Python and connected to the simulator via an ethernet cable. The loop would stop after five iterations. GUI = graphical user interface.

There were five levels of difficulty in the VR training system as shown in Fig. 2. In consultation with a senior flight instructor and the flight simulator developers, (degraded) visibility was chosen as the primary factor to adjust flight task difficulty and manipulate pilot workload. In levels 1 through 3, outside visuals gradually became less clear due to increasingly dense cloud formations. To increase workload further, level 4 introduced a modification to the flight instruments by disabling the attitude indicator, thereby removing the artificial horizon from the virtual cockpit.

The highest difficulty level, level 5, featured a false horizon created by a sloping cloud formation. This condition was particularly challenging because sloping clouds are often mistaken for the true horizon. This misconception can lead pilots to align the (virtual) aircraft with the cloud line, resulting in disorientation and diminished performance [28].

## II. Experiment

### A. Participants

Fifteen Royal Netherlands Air Force pilots participated in this study (2 females, 13 males; $M_{age}$ = 24.50 years, $SD_{age}$ = 1.70 years). The average flight experience of the sample was 62.60 hours (*SD* = 91.60, *Median* = 35, range: 0-380). The pilots completed an average of 14.40 flight hours (*SD* = 14.80, *Median* = 20, range: 0-40) over the last 90 days prior to participation. At the beginning of the experiment, the pilots gave their informed consent.

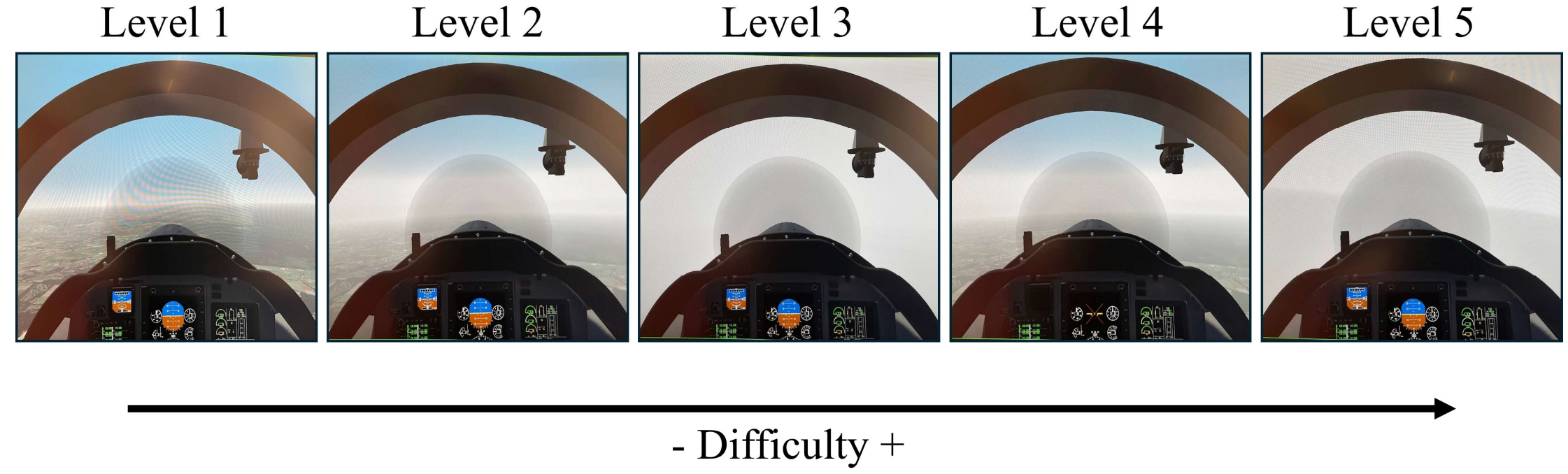

**Fig 2.** The five levels of difficulty employed in the current study. Level 1: clear weather and outside visuals, Level 2: misty outside visuals, Level 3: heavy fog resulting in no outside visuals, Level 4: misty outside visuals and failure of artificial horizons within cockpit instruments, Level 5: heavy fog with false horizon.

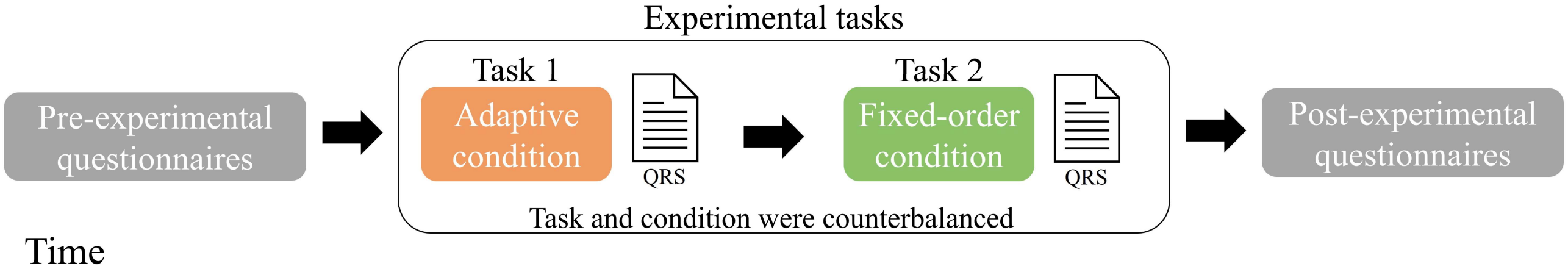

**Fig. 3.** Experimental design. All subjects performed one task in the Adaptive condition, and another task in the Fixed-order condition. The order of the two tasks and the two conditions was counterbalanced. QRS = questionnaires.

*B. Experimental procedure*

The experimental procedure is presented in Fig. 3. The experiment included two conditions: 1) Fixed-order condition in which levels of difficulty increased incrementally from easy (level 1) to difficult (level 5), and 2) Adaptive condition in which the levels of difficulty increased or decreased based on workload classifications as explained previously in section II. B. *Real-time implementation*, with the first trial beginning at the medium difficulty level (level 3), and completing after a total of five trials.

Two different flight tasks were employed to prevent a learning effect over conditions, namely (1) deceleration; changing the aircraft speed from 180 kts to 110 kts while maintaining straight and level, and (2) medium turn; a full 360° turn to the left on a roll angle of 30° while maintaining altitude. These tasks were selected from the PC-7 training curriculum and in consultation with a flight instructor. Conditions and tasks were counterbalanced across subjects. The pilots were not made aware of the difference between the two conditions until after the experiment.

*C. Materials*

The training was conducted in a VR flight simulator of a fixed-wing PC-7 Turbo Trainer aircraft (multiSIM B.V., the Netherlands), featuring a cockpit mock-up including stick and pedals with control loading (Fig. 4). The VR environment was displayed using the Varjo Aero headset (Varjo Technologies Oy, Finland). Flight parameters were recorded during the tasks with a sample rate varying up to 500 Hz. Brain signals were recorded with a wireless 32-channel EEG system at 250 Hz (g.Nautilus PRO, g.tec medical engineering GmbH, Austria). Electrodes were placed according to the 10-20 system, see [25] for an overview of the channels and placement.

The experimental questionnaires included a pre-experimental questionnaire inquiring about the pilots' demographics, flight experience and experience with VR. Each experimental condition was concluded with the (raw) NASA Task Load Index (NASA-TLX; [29]), Simulator Sickness Questionnaire (SSQ; [30]) and User Engagement Scale short form (UES-SF; [31]). After every single trial within the conditions, pilot workload was inquired using the Instantaneous Self-Assessment of workload technique (ISA; [32], [33]), as well as their mental fatigue using the Fatigue Instantaneous Self-Assessment (F-ISA; [34]). Both the ISA and F-ISA are rated on a 5-point Likert scale.

All experiments ended with a semi-structured interview inquiring about the pilots' experience with the training system. The interview focused on topics such as perceived difficulty, task engagement, and whether participants noticed any differences between the conditions. Specifically, it included the following questions: (1) did you notice a difference between the two training sessions? If yes, what difference did you notice? (2) Which training did you prefer? (3) Which training session was more engaging? and (4) Which training session was more difficult? In-between questions (1) and (2) of the semi-structured interviews, pilots were informed of the experimental conditions and encouraged to verbalize their thoughts. Additional commentary was also encouraged throughout the interviews and transcribed.

*D. Data analysis*

The data processing and analysis were performed in MATLAB R2023b [35], and RStudio [36] with R version 4.3.3. [37].

*Flight performance*

Flight performance was estimated using the same approach as in [38]: Root Mean Square Errors (RMSEs) from target parameters were calculated for the pitch and roll axes of the virtual aircraft for each trial. The function "findchangepts" of the Signal Processing Toolbox [35] was used to locate points in time that pilots initiated a roll during medium turn tasks, and where applicable, its completion. The RMSEs of pitch and roll were summed to create one metric of flight performance. A larger (summed) RMSE value indicates larger deviations from target parameters, hence a decreased performance.

*Statistical tests*

The data were checked for outliers and normality. One extreme value (greater than three times the interquartile range from the median) was identified. However, this datapoint was not influential as its removal did not affect our statistical results. Subsequently, it was retained. The normally-distributed data included the NASA-TLX scores for subjective workload, UES-

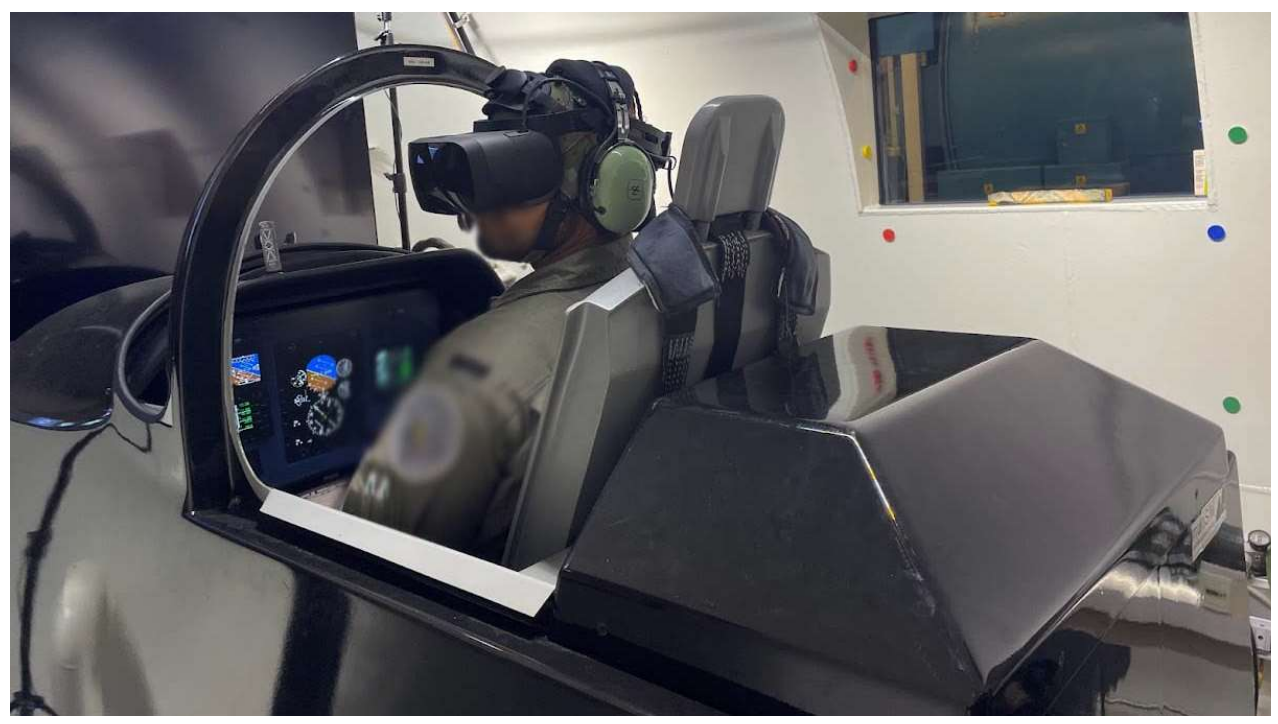

**Fig 4.** Experimental set-up in mock-pit with head-mounted display, cockpit instruments, headphones and EEG cap.

SF scores for subjective engagement, and flight performance. Non-parametric data included the SSQ scores for simulator sickness.

Repeated-measures ANOVAs were conducted to examine the effects of Condition (Fixed-order vs. Adaptive) and Task (deceleration vs. medium turn), as well as their interaction, on each outcome measure (i.e., NASA-TLX scores for subjective workload, SSQ scores for simulator sickness, UES-SF scores for subjective engagement, and flight performance). Where a main effect was observed, post-hoc comparisons were conducted with paired *t*-tests or Wilcoxon signed-rank tests as appropriate.

We also conducted two Spearman's correlation tests between flight performance (continuous data) and ISA scores for workload (ordinal data), one for each condition. Additionally, we performed two Spearman's correlation tests for flight performance and F-ISA scores for fatigue (ordinal data), again one for each condition. The Spearman's method is based on ranks and thus is appropriate for ordinal data [39], and is not sensitive to outliers [40]. To determine whether the Spearman's correlation coefficients differed significantly between Fixed and Adaptive conditions, we employed Fisher's *Z*-tests. This method compares correlation coefficients by transforming them into Fisher's *Z*-scores and then assesses the difference between them.

In all statistical tests (repeated-measures ANOVA tests, correlation tests and Fisher's *Z*-tests), False Discovery Rate (FDR) adjustments were applied to the *p*-values to reduce inflation of type I error due to multiple testing [41].

## E. RESULTS

### *A. Subjective workload (NASA-TLX)*

The ANOVA of the NASA-TLX scores revealed no significant interaction between Condition and Task, $F(1, 13)$ =

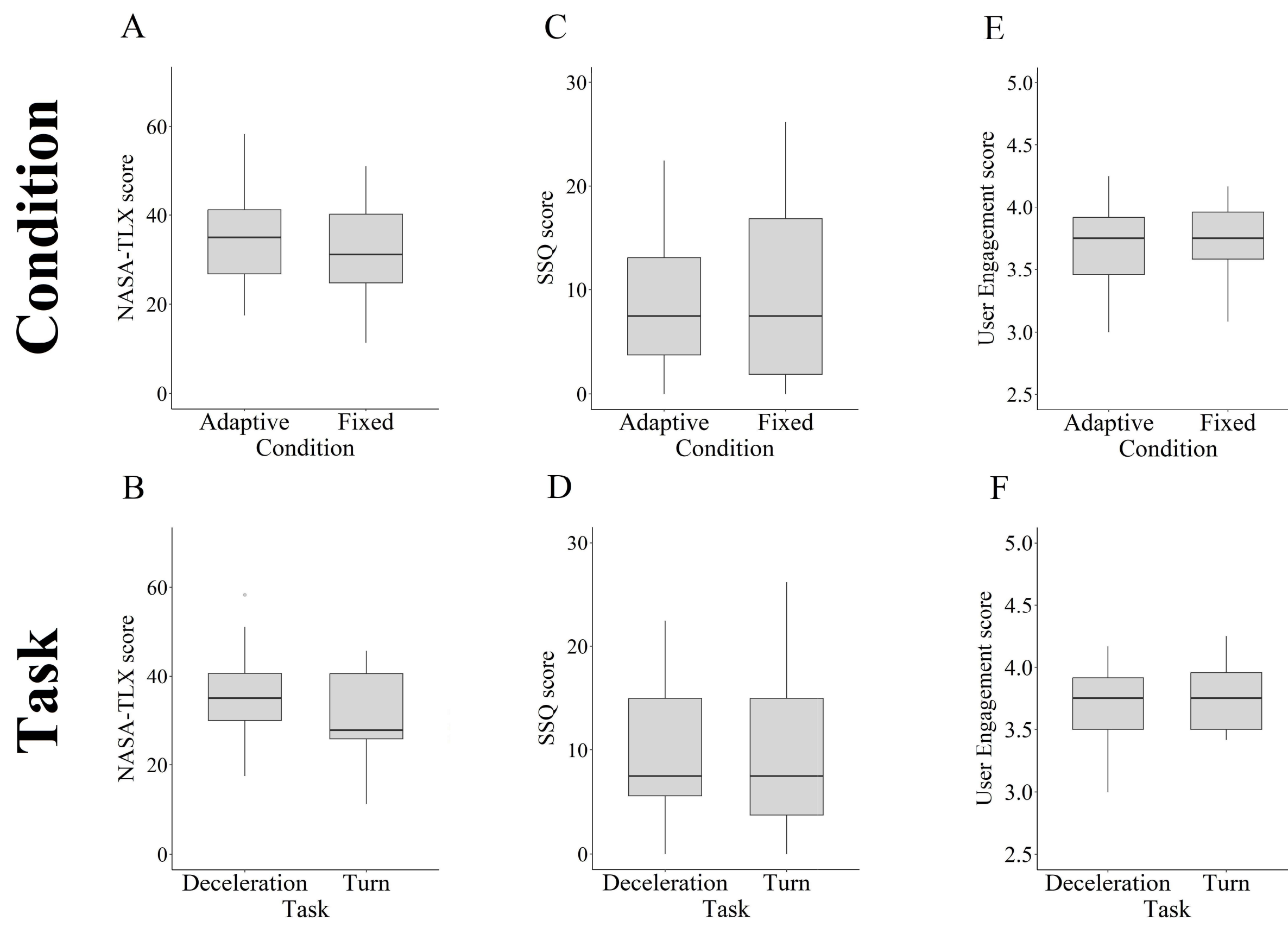

**Fig. 5.** Subjective scores for workload (A and B), simulator sickness (C and D) and user engagement (E and F).

1.82, $p$ = .48, $\eta_p^2$ = .12, 95% CI [.00, .46], indicating that the effect of Condition on subjective workload did not depend on the flight task. Additionally, no main effect was observed for Task ($F$(1, 13) = 5.47, $p$ = .16, $\eta_p^2$ = .30, 95% CI [.00, .60]), nor for Condition ($F$(1, 13) = 1.72, $p$ = .48, $\eta_p^2$ = .12, 95% CI [.00, .46]). See Fig. 5 A and B for the boxplots of NASA-TLX scores per condition and per task.

*B. Simulator sickness questionnaire (SSQ)*

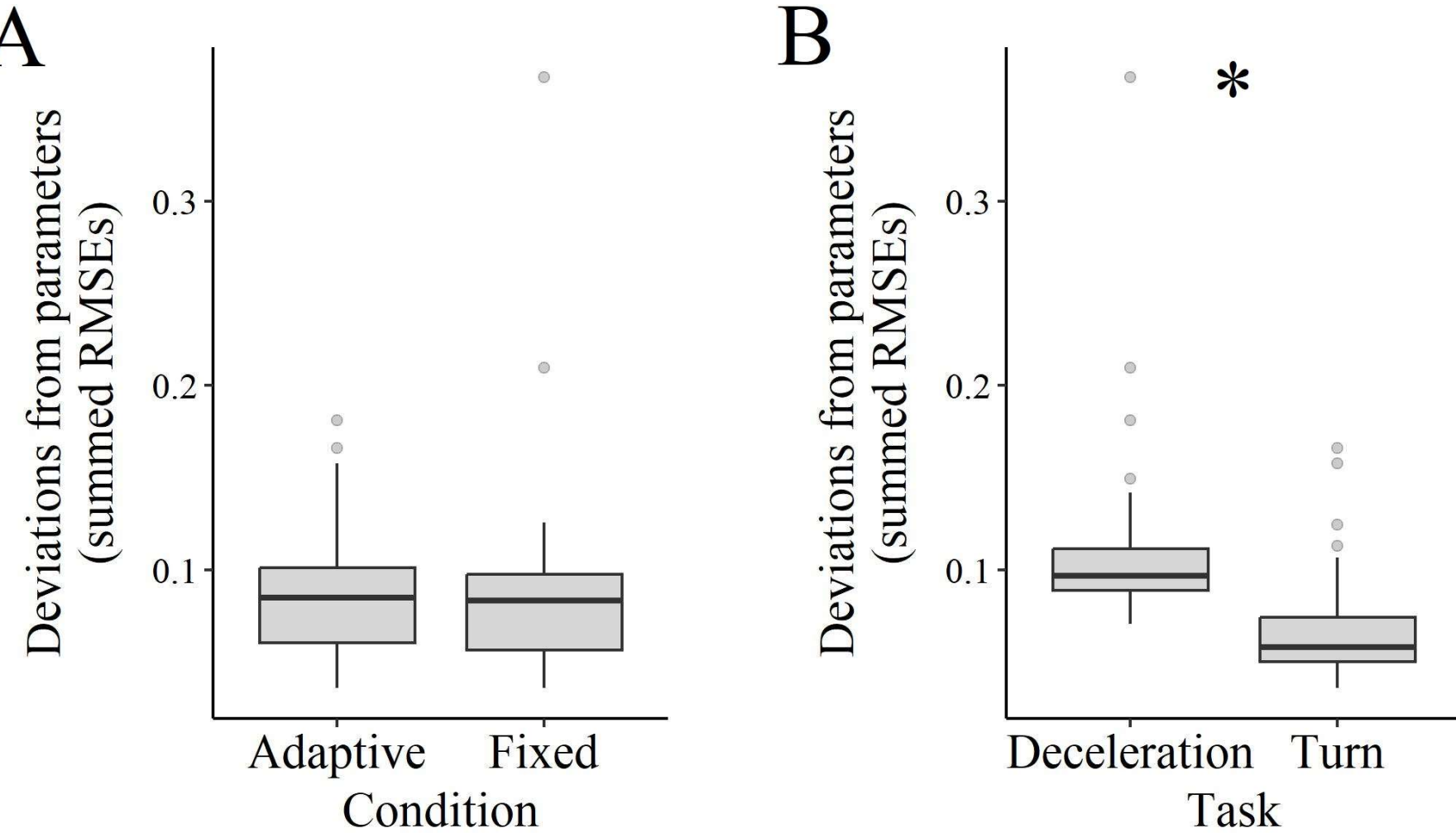

**Fig. 6.** Boxplots of flight performance per condition (A) and task (B). * indicates a statistically significant difference of $p < .001$.

The analysis revealed no significant interaction between Condition and Task ($F$(1, 13) = 1.43, $p$ = .48, $\eta_p^2$ = .10, 95% CI [.00, .44]) on the SSQ, suggesting that the effect of Condition on SSQ scores did not depend on the task performed.

Both the main effects of Condition ($F$(1, 13) = .09, $p$ = .83, $\eta_p^2$ = .01, 95% CI [.00, .26]) and Task ($F$(1, 13) = .08, $p$ = .83, $\eta_p^2$ = .01, 95% CI [.00, .26]) were also non-significant, indicating that neither Condition nor Task had a statistically significant impact on SSQ scores. See Fig. 5 C and D for the boxplots of SSQ scores per condition and per task.

*C. User Engagement Scale (UES-SF)*

The ANOVA results indicate that neither the interaction between Condition and Task ($F$(1, 13) = .17, $p$ = .83, $\eta_p^2$ = .01, 95% CI [.00, .29]) nor the main effects of Condition ($F$(1, 13) = .17, $p$ = .83, $\eta_p^2$ = .01, 95% CI [.00, .29]) or Task ($F$(1, 13) = 1.63, $p$ = .48, $\eta_p^2$ = .11, 95% CI [.00, .45]) were statistically significant. These findings suggest that changes in the condition or type of task do not significantly influence user engagement scores. See Fig. 5 E and F for the boxplots of UES-SF scores per condition and per task.

*D. Flight performance*

The results from the ANOVA indicated that there were no significant interaction effects between Condition and Task on pilots' flight performance ($F$(1, 13) = 1.35, $p$ = .48, $\eta_p^2$ = .09, 95% CI [.00, .43]).

The main effect of Condition was also not significant ($F$(1, 13) = .16, $p$ = .83, $\eta_p^2$ = .01, 95% CI [.00, .29]). However, the main effect of Task was significant ($F$(1, 13) = 116.29, $p$ < .0001, $\eta_p^2$ = .90, 95% CI [.75, .95]), indicating that pilots' performance differed across the tasks.

The post-hoc comparison test confirmed a significant difference in performance between the deceleration task ($M$ = .10, $SD$ = .01) and medium turn ($M$ = .06, $SD$ = .01), $p$ < .0001, see Fig. 6.

*E. Correlation flight performance and subjective workload (ISA)*

We observed a strong positive relationship between subjective workload (ISA scores) and deviations from target parameters for the Fixed condition, $\rho$ = .56, $p$ < .001, as well as for the Adaptive condition, $\rho$ = .30, $p$ < .05. These results confirmed that, as expected, an increase in pilot's workload was associated with increased deviations from flight target parameters increased, i.e., decrease in pilots' performance (Fig. 7).

Fisher's $Z$-test for the comparison of the two correlations revealed a $Z$-score difference of 1.95, with $p$ = .18.

*F. Correlation flight performance and subjective fatigue (F-ISA)*

As shown in Fig. 8, there were no significant correlations between subjective fatigue (F-ISA scores) and deviations from target parameters in neither the Fixed-order condition ($\rho$ = .08, $p$ = .76), nor the Adaptive condition ($\rho$ = .09, $p$ = .75). Fisher's $Z$-test for the comparison of the two correlations revealed a $Z$-score difference of -.06, with $p$ = .96, suggesting no difference between the correlation coefficients.

*G. Workload classifications*

Table I and Table II of the Supplementary Materials [42] display the EEG-based workload classification outputs of the system for each of the trials, per subject, within the Fixed-order condition and Adaptive condition, respectively. The outputs offer insights into the model's performance under the varying difficulty levels and the two conditions, highlighting where it may have successfully differentiated workload levels and where it may have encountered limitations in doing so based on the real-time EEG signals. From the tables, it appears that there is no clear pattern in the trials and conditions across the pilots as there is a high variation in classifier outputs. Noteworthy, pilots 1, 6, 11, 12 and 14 showed no variation in classifier outputs, regardless of condition or difficulty level.

### *H. Semi-structured interviews*

Each of the experiments was concluded with a semi-structured interview. Results showed that none of the subjects explicitly noticed that the Adaptive scenario was personalized according to their workload. Ten of the subjects noticed the difference in the order of the difficulty levels between the two conditions, but five subjects did not. There were individual differences in the perceived difficulty of the two tasks at different visibility levels. Some pilots mentioned being bored on occasion, while being engaged and focused at other times during the experiment.

Once the experimenter explained the difference between the two training conditions, pilots were asked to choose the more difficult condition and their overall preference. Ten of the 15 pilots preferred the Adaptive training over the Fixed-order training, and nine of the 15 subjects found that the Adaptive condition was more difficult. According to one of the pilots: “The adaptive version has my preference, because with a fixed order you can expect what comes next and prepare for that. With the adaptive one, you do not know what’s to come and that is more realistic and in line with the experience in the (real) aircraft.”

One of the pilots provided a general comment about the simulation training: “it is more fun when the workload is somewhat higher”. Another mentioned, after informing them about the Adaptive condition: “I can recognize that the simulator adapted to me, because the start was hard, then it became easier, and then it became harder again”. For another pilot, the experience was the opposite for the Adaptive condition: “the first one (*author’s note:* referring to the Adaptive condition) stayed or became easy”. Finally, one of the pilots mentioned a slight discomfort with wearing the VR headset, EEG cap and headphones altogether.

## IV. Discussion

In this study, we created a neuro-adaptive VR flight training system for adjustment of difficulty level in flight tasks using a pre-trained ML model for classifying workload (model previously reported in [25]). We evaluated the adaptive flight training system by comparing it to a fixed-order system in an experiment with novice military pilots. We compared subjective workload, engagement, simulator sickness, and flight performance. Semi-structured interviews were held at the end of the experiment to explore the overall user experience with the neuro-adaptive VR flight training system and flight tasks.

The analyses did not reveal any significant differences between the Adaptive and Fixed-order conditions, nor any interactions between the conditions and the type of task across all outcome variables, i.e., the subjective measures of workload, engagement, and simulator sickness, as well as the objective metric for flight performance. The semi-structured interviews revealed that while pilots did not consciously notice a difference between the two experimental conditions, the majority expressed a preference for the Adaptive condition when the difference was communicated. An explanation could be the variability and unpredictability in difficulty levels in the Adaptive condition, as noted by some participants and supported by the workload classification outputs. However, this was very subject-dependent: five pilots experienced the Adaptive condition remaining at the same difficulty level throughout multiple trials due to stable classification outputs. It is possible that these contradicting experiences within the sample have cancelled out any possible effects for subjective engagement and workload. Alternatively, a ceiling effect of these relatively experienced participants ($M$ = 62.60 flight hours, *Median* = 35 flight hours) cannot be ruled out. Furthermore, albeit not exceptionally different from sample sizes in comparative studies, the relatively small and homogeneous sample of novice military pilots may have limited the reliability of these findings, and their generalizability to broader and/or different pilot populations..

Next, we found significant correlations between flight performance and subjective workload in both conditions of the experimental study. This relationship is quite common in the literature, e.g., [42], [44], and shows that the utilized VR training system and chosen tasks are capable of inducing variability in pilots’ experienced workload. We did not observe any relationship between flight performance and fatigue, possibly because the duration of experiments was not long enough to induce significant pilot fatigue. There were also no differences in the strength of these relationships between the conditions.

An explanation for the null-results regarding the lack of observed differences between experimental conditions could be the limitations of our subject-independent and calibration-free approach.

Also of note is that our study faced a double transfer learning problem. First, the ML model was trained on data from a previous group of novice pilots and then applied to a new group of pilots. A similar issue was observed in previous literature [45], where an attempt was made to identify a subject-independent neural indicator for task proficiency in a navigation task displayed on a desktop monitor. Their classifier was tested and evaluated per subject, but was pre-trained on EEG data from the remaining participants. Although the classifier performed well for nine out of 15 participants, it showed limited success for the others in the sample. These findings are consistent with the current state-of-art, which suggests that subject-independent BCIs do not perform equally well for all individuals due to innate differences in neural activity or anatomy, i.e., BCI inefficacy [46], [47].

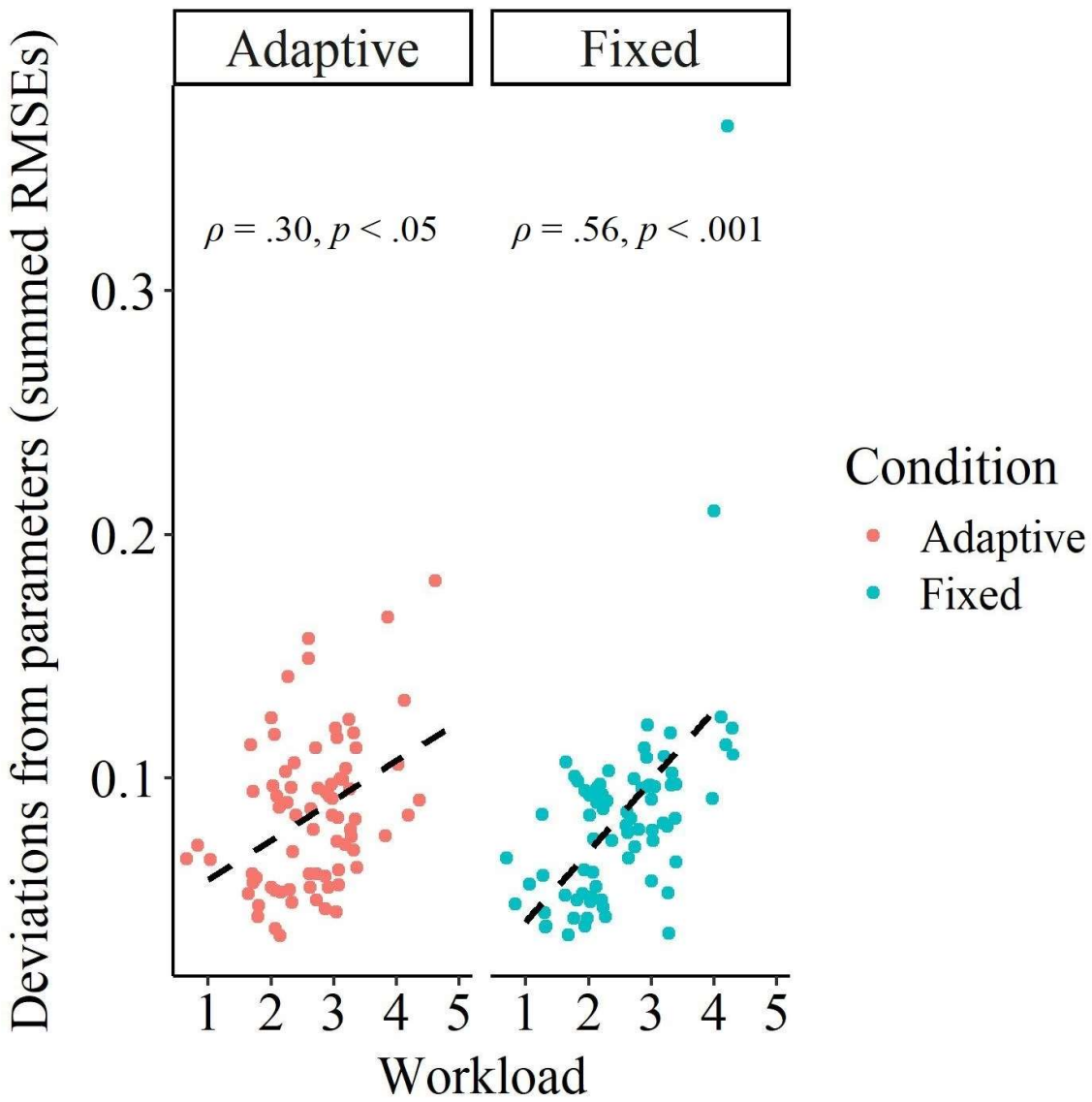

**Fig. 7.** Scatter plot of flight performance (deviations from target parameters) and subjective workload. Dashed lines display the correlations per condition.

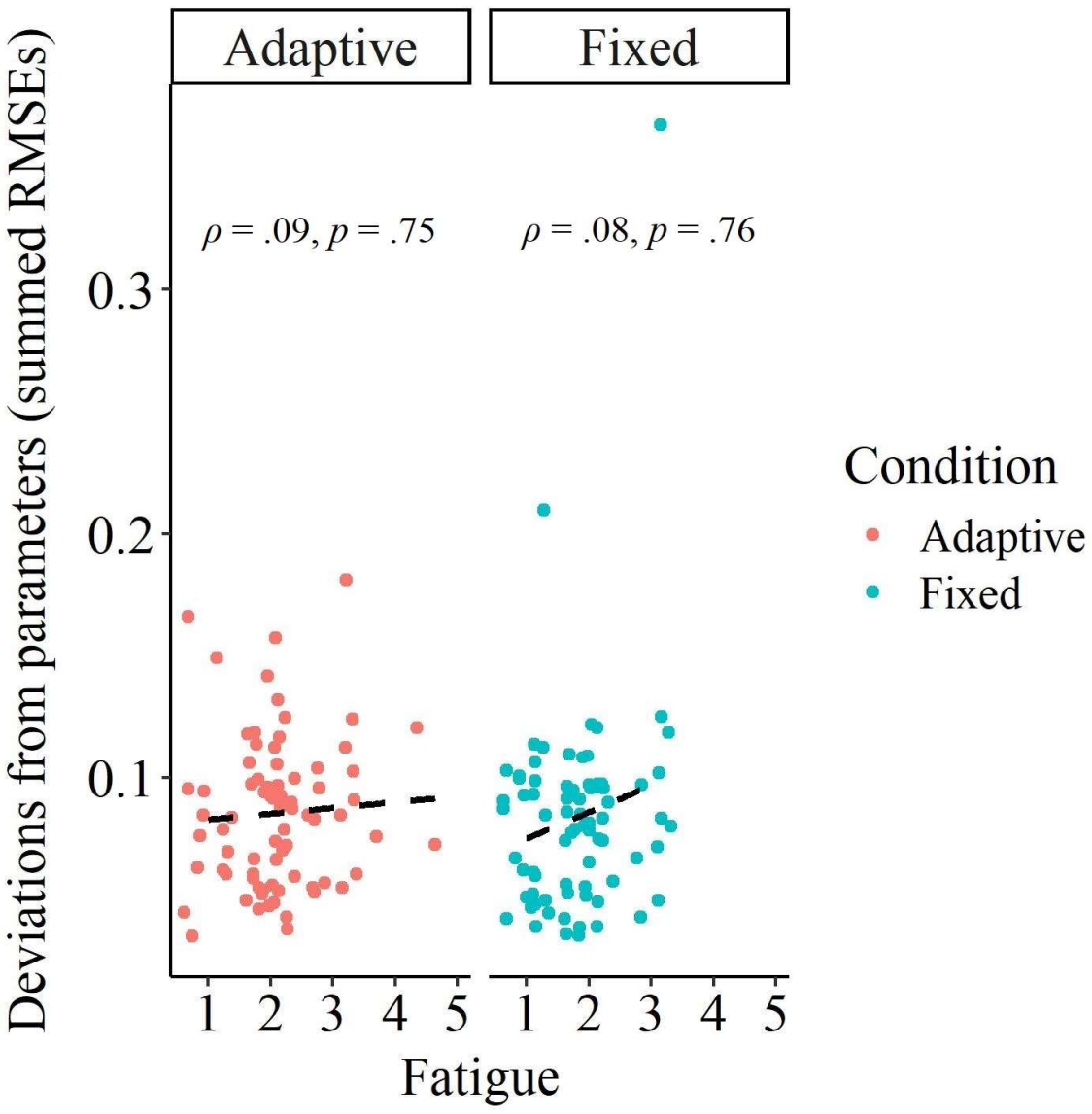

**Fig. 8.** Scatter plot of flight performance (deviations from target parameters) and subjective fatigue. Dashed lines display the correlations per condition.

Second, the tasks in the original study were different from the tasks in de current study. The ML model was trained on data involving a deceleration task and an auditory N-back for workload manipulation. In the current study, we employed two fight tasks (a deceleration task and a medium turn task) to prevent a learning effect between conditions. Additionally, the difficulty of the tasks was manipulated in a different manner: workload was manipulated by degrading the visibility of the environment in the VR simulation. According to Wickens' [48] multiple-resource theory, tasks can engage different cognitive and perceptual resources separately, such as visual vs. auditory. Assuming this theory is valid, it is possible that our pre-trained ML model is designed to identify auditory load or the presence/absence of dual-task processing, and not the overall workload that we were aiming to identify.

A technical limitation in the current study was the failure of Artifact Subspace Reconstruction (ASR) within the EEG pre-processing pipeline. However, this should have not posed a problem, since ICA is sufficient for EEG data cleaning [49].

The transfer learning problem may be mitigated by a short calibration of the pBCI for each pilot. For instance, in a case study (of two co-pilots) for real-time workload and arousal monitoring, Borghini et al. [13] recorded "eyes-closed" EEG calibration data from each pilot to find their individual's alpha peaks and to compute the individual alpha frequency (IAF). The IAF was then used to define theta and alpha frequency bands per pilot. This approach was applied in simulated flight missions, and seemed very promising.

For future research, it would be beneficial to implement a sham-adaptive order of the difficulty levels (i.e., an order that mimics a neuro-adaptive order, but is actually fixed or randomized) as a control condition to eliminate potential effects of participants' expectations. Future work could also further expand the current study by considering a multi-modal approach for workload classification, for instance by adding the performance metric as a feature – only when target parameters are identified and known to pilots. This multi-modal approach is promising, especially because of the linear relationship that was observed between flight performance and subjective workload in the current study. A three-class approach with low (underload), intermediate, and high workload (overload) may also be more effective than the current two-class model. Prior studies [50] reached a multiclass classification accuracy of .83 utilizing VR simulated aircraft operations to induce different levels of workload.

### *A. Recommendations*

During the prototyping and evaluation of our real-time adaptive VR flight training system, we identified several challenges and pitfalls. Therefore, to advance future developments of real-time pBCIs in flight training, we share some recommendations for future work:

1) *Calibration:* As previously discussed, a simple and brief calibration phase, such as the implementation of IAF, could finetune both subject-dependent and -independent ML models and predictions.
2) *Task-selection:* Given that transfer learning across subjects and tasks remains a challenge in pBCIs, it is recommended to identify the final application of the pBCI, and select eligible users and tasks for collecting the ML training data accordingly.
3) *Human-in-the-loop:* When prototyping and evaluating a pBCI for real-time usage, one should start with human-in-the-loop pilot studies to identify areas for improvement and facilitate informed decision-making during flight training sessions. Automated decisions made by the system should be carefully interpreted and, when necessary, corrected by the human experimenter to optimize the system's design and accuracy.

## V. Conclusion

In this study, we presented and evaluated a prototype of a neuro-adaptive VR flight training system, tested in real-time with novice military pilots. While quantitative measures revealed no significant changes in engagement, workload, or flight performance over a fixed-order system, the consistent relationship between workload and performance reinforces the importance of adaptive approaches in managing training demands. Importantly, qualitative feedback highlighted a clear preference for the adaptive system, with participants valuing the variability it introduced into the training sessions. Future work should refine adaptive mechanisms and evaluate longer training programs to fully realize the potential of neuro-adaptive systems in high-demand training environments. Findings of the current study suggest that neuro-adaptive VR training may not immediately enhance measurable performance outcomes, yet still holds promise for improving pilot motivation, training sustainability, and long-term skill acquisition.

# Supplementary Materials

*Table I.* Difficulty level and workload classification per trial within the Fixed-order condition.

| **Subject** | **Trial 1** | | **Trial 2** | | **Trial 3** | | **Trial 4** | | **Trial 5** | |
|---|---|---|---|---|---|---|---|---|---|---|
| | **Difficulty** | **Output** | **Difficulty** | **Output** | **Difficulty** | **Output** | **Difficulty** | **Output** | **Difficulty** | **Output** |
| 1 | 1 | Low | 2 | Low | 3 | Low | 4 | Low | 5 | Low |
| 2 | 1 | High | 2 | Low | 3 | Low | 4 | Low | 5 | Low |
| 3 | 1 | High | 2 | High | 3 | High | 4 | High | 5 | Low |
| 4 | 1 | High | 2 | Low | 3 | Low | 4 | High | 5 | High |
| 5 | 1 | High | 2 | High | 3 | High | 4 | High | 5 | High |
| 6 | 1 | Low | 2 | Low | 3 | Low | 4 | Low | 5 | Low |
| 7 | 1 | High | 2 | High | 3 | High | 4 | High | 5 | Low |
| 8 | 1 | Low | 2 | Low | 3 | Low | 4 | Low | 5 | Low |
| 9 | 1 | Low | 2 | Low | 3 | Low | 4 | High | 5 | High |
| 10 | 1 | High | 2 | Low | 3 | High | 4 | High | 5 | High |
| 11 | 1 | Low | 2 | Low | 3 | Low | 4 | Low | 5 | Low |
| 12 | 1 | High | 2 | High | 3 | High | 4 | High | 5 | High |
| 13 | 1 | Low | 2 | Low | 3 | High | 4 | Low | 5 | Low |
| 14 | 1 | Low | 2 | Low | 3 | Low | 4 | Low | 5 | Low |
| 15 | 1 | Low | 2 | Low | 3 | Low | 4 | Low | 5 | Low |

*Note.* Output = output of the workload classifier, Low = low workload, High = high workload.

*Table II.* Difficulty level and workload classification per trial within the Adaptive condition.

| **Subject** | **Trial 1** | | **Trial 2** | | **Trial 3** | | **Trial 4** | | **Trial 5** | |
|---|---|---|---|---|---|---|---|---|---|---|
| | **Difficulty** | **Output** | **Difficulty** | **Output** | **Difficulty** | **Output** | **Difficulty** | **Output** | **Difficulty** | **Output** |
| 1 | 3 | Low | 4 | Low | 5 | Low | 5 | Low | 5 | Low |
| 2 | 3 | High | 2 | Low | 3 | High | 2 | High | 1 | Low |
| 3 | 3 | Low | 4 | Low | 5 | Low | 5 | Low | 5 | High |
| 4 | 3 | Low | 4 | High | 3 | High | 2 | Low | 3 | High |
| 5 | 3 | High | 2 | High | 1 | Low | 2 | High | 1 | High |
| 6 | 3 | Low | 4 | Low | 5 | Low | 5 | Low | 5 | Low |
| 7 | 3 | High | 2 | High | 1 | High | 1 | High | 1 | High |
| 8 | 3 | Low | 4 | Low | 5 | Low | 5 | High | 4 | Low |
| 9 | 3 | Low | 4 | Low | 5 | High | 4 | Low | 5 | High |
| 10 | 3 | Low | 4 | Low | 5 | Low | 5 | Low | 5 | Low |
| 11 | 3 | Low | 4 | Low | 5 | Low | 5 | Low | 5 | Low |
| 12 | 3 | High | 2 | High | 1 | High | 1 | High | 1 | High |
| 13 | 3 | Low | 4 | High | 3 | Low | 4 | Low | 5 | Low |
| 14 | 3 | Low | 4 | Low | 5 | Low | 5 | Low | 5 | Low |
| 15 | 3 | Low | 4 | Low | 5 | High | 4 | Low | 5 | Low |

*Note.* Output = output of the workload classifier, Low = low workload, High = high workload.